\documentclass[journal]{IEEEtran}
\ifCLASSINFOpdf
\else
\fi
\usepackage{graphicx}
\usepackage{caption}
\graphicspath{{./images/}}
\begin{document}
%

\title{Vulnerability Detection in Open Source Software: An Introduction}
%
%
%

\author{Stuart Millar, Rapid7 LLC, \textit{stuart\_millar@rapid7.com}

\thanks{S. Millar is with Rapid 7 LLC, Boston, MA, USA.  This version dated March 26th 2017.  For correspondence please e-mail: stuart\_millar@rapid7.com}
}

\maketitle


\begin{abstract}
This paper is an introductory discussion on the cause of open source software vulnerabilities, their importance in the cybersecurity ecosystem, and a selection of detection methods.  A recent application security report showed 44\% of applications contain critical vulnerabilities in an open source component, a concerning proportion.  Most companies do not have a reliable way of being directly and promptly notified when zero-day vulnerabilities are found and then when patches are made available.  This means attack vectors in open source exist longer than necessary.  Conventional approaches to vulnerability detection are outlined alongside some newer research trends.  A conclusion is made that it may not be possible to entirely replace expert human inspection of open source software, although it can be effectively augmented with techniques such as machine learning, IDE plug-ins and repository linking to make implementation and review less time intensive.  Underpinning any technological advances should be better knowledge at the human level.  Development teams need trained, coached and improved so they can implement open source more securely, know what vulnerabilities to look for and how to handle them.  It is the use of this blended approach to detection which is key.
\end{abstract}


%
\IEEEpeerreviewmaketitle

\section{Introduction}
%
%
%
%

Open source software (OSS) is developed collaboratively in the public domain with a licence granting rights to the user base that are usually reserved for copyright holders.  A well-known OSS licence is the GNU General Public Licence that allows free distribution under the condition that further developments are also free.  In a globally connected software society, a sizeable amount of development work is effectively crowd-sourced to an international community of OSS developers, with little awareness of the potential security problems this creates \cite{13tools}. OSS libraries increase development speed but there is a tangible increase in risk also, with the Heartbleed bug in OpenSSL being a prime example.  Research into vulnerability detection in OSS is crucial given its prevalence with many companies using vulnerable OSS components and vulnerable libraries being repackaged in software.  This OSS uptake shows no sign of reversing or slowing, with a recent survey \cite{blackduck} indicating that 43\% of respondents think OSS is superior to its commercial equivalent.
 
At the crux of OSS vulnerability is that today’s applications commonly use thirty or more libraries which in turn can comprise up to 80\% of the code in any such application.  These libraries have the same full privileges of the application that use them, letting them access data, write to files or send data to the internet.  Anything the application can do, the library can do.  Some estimate that custom built Java applications contain 5-10 vulnerabilities per 10,000 lines of code \cite{williams}.  A library can have on average 10,000 to 200,000 lines of code, therefore the chances a library has never had a vulnerability are very slim, with it perhaps being more likely in fact that it has not even been examined for vulnerabilities.  Hence libraries with no vulnerabilities should not automatically be considered ‘safe’.  Most vulnerabilities are undiscovered and it might be said the only way to deal with the risk of unknown vulnerabilities is to have someone who understands security manually analyse the source code.  Tool support provides hints but is not a replacement for experts because the lack of context within libraries makes it very difficult for current tools at time of writing to conclusively identify vulnerabilities.

This paper contributes a short introductory discussion on vulnerability detection in OSS, and is organised as follows:  Section II is a background overview and Section III contains information on conventional and emerging detection methods.  Conclusions are presented with ideas for future work in Section IV.


\section{Background}

A study carried out in 2012 \cite{williams} found that more than 50\% of the Fortune Global 500 companies have downloaded vulnerable OSS components, security libraries and web frameworks.  This report analysed 113 million Java framework and security library downloads by more than 60,000 commercial, government and non-profit organisations from the Central Repository.  Central is the software industries most widely used repository of OSS with more than 300,000 libraries.  It was found the vast majority of library flaws remain undiscovered, the presence of a vulnerability (or an absence of one) is not a security indicator, and that typical Java applications are likely to include at least one vulnerable library.  Furthermore, the same study showed most organisations did not have a strong process in place for ensuring the libraries they rely upon are up-to-date and free from vulnerabilities.  The authors of \cite{williams} stress there are no shortcuts and they go as far as saying the only useful indicator of library security is a thorough review that finds minimal vulnerabilities – in other words, software assurance, or the measure of how safe the software is to use, needs to be generated internally.  One might say this is surprising, as in many other product or service industries, this assurance – consider it some kind of warranty or seal of approval perhaps – is offered up by the supplier without hesitation to help build trust and sell to the customer.  \cite{pham} adopt a similar stance, agreeing that recurring vulnerabilities in software are due to reuse.  This reuse includes the same code base with an identical or very similar code structure, method calls and variables.  Interestingly these attributes form the basis of a proposed method of detecting unreported vulnerabilities in one system by consulting knowledge of reported vulnerabilities in other systems that reuse the same code.

Linus’ Law \cite{linus} is often quoted in relation to OSS, which is “given enough eyeballs, all bugs are shallow”, meaning with a large enough number of developers looking at code, errors can be found.  However, this can questioned from a scientific viewpoint, and an empirical study of Linus’ Law appeared to show more collaboration meant more vulnerabilities, not less.  \cite{linusoss} found that files with changes from nine or more developers were sixteen times more likely to have a vulnerability than files changed by fewer than nine developers.  Thus inherent collaborative nature of OSS creates potentially unavoidable vulnerabilities that require addressing.

\section{OSS Vulnerability Detection}

\subsection{Conventional Detection Methods}
\label{subsec:methodology}

There are not currently an abundance of publications on vulnerability detection in OSS.  However, those that have been written thus far describe three conventional methods – static analysis, dynamic analysis and code reviews. 

\subsubsection{Static Analysis}

Many black-box static analysis techniques and tools scan source code and detect vulnerabilities in software after it has been written, which encourages late detection and produces a lot of false positives .  \cite{SAMPAIO2016337} explicitly referenced the cut and thrust of the software development process, saying that external static tools for secure programming don’t fit into such a workflow, since they don’t work with the IDE and are retrospective.  \cite{ZHANG201460} concur that static analysis produces high levels of false positives, as do \cite{Grieco} and \cite{Perl}.  \cite{Shahmeri} point out it is hard to know which vulnerabilities a static analysis tool deals with, and there are difficulties in obtaining and maintaining up-to-date tooling.  

\cite{GOSEVAPOPSTOJANOVA201518} specifically wrote about the capability of static code analysis to detect vulnerabilities, concluding that tools are not effective.  They tested three widely used commercial tools and found 27\% of C/C++ vulnerabilities and 11\% of Java vulnerabilities in their dataset were missed by all three.  In some cases, they were comparable to or worse than random guessing.  They too make the point about tools being prone to false positives, and this consolidates the need to find other methods of detection rather than rely solely on static analysis.  That is not to say static analysis is of little use, as some compliance regulations require inventories of OSS components so that risks can be addressed.  Static tools can scan open source code and create an inventory, so when a new vulnerability is disclosed, it is known which applications use the vulnerable OSS.  The OWASP Dependency-Check tool \cite{owaspcheck} analyses code and creates reports on associated CVE entries.

\subsubsection{Dynamic Analysis}

White-box dynamic analysis can also be called run-time analysis.  Fuzzing is often used here, where inputs are changed using random values to detect unwanted behavior \cite{Shahmeri}.  \cite{hafizgame} researched the nuances of how vulnerabilities were discovered by researchers, and how those same researchers shared their findings with the OSS community.  They found running a fuzzer and debugging was the chosen method for developers exploring binary executables to find buffer overflows.  Vulnerability researchers tend to make their own fuzzing tools, seeing it as part of the learning process and preferring this approach over more systematic exploration methods.  \cite{Grieco} notes the usefulness of fuzzing, and that it needs only basic knowledge to undertake. However fuzzing does not allow the control of program execution, large campaigns are needed for results, and it is time consuming.  \cite{ZHANG201460} contends fuzzing doesn’t scale if dynamic symbolic execution is used, as it explores code paths simultaneously which could create large workloads.  Symbolic execution uses symbolic values for variables instead of concrete values to execute all paths in a program.

\subsubsection{Manual Code Reviews}

These involve manual inspection of the source code in a white-box manner.  Consequently, this method requires a lot of human effort \cite{Perl}.  Working on source code manually does however detect vulnerabilities \cite{hafizgame}, and recall that \cite{williams} argued code reviews, conducted by someone with appropriate security knowledge, may be the only way to properly deal with vulnerabilities.

\begin{figure}[b]
    \centering
    \captionsetup{justification=centering}
    \captionsetup{font=small}
    \includegraphics[width=0.8\columnwidth]{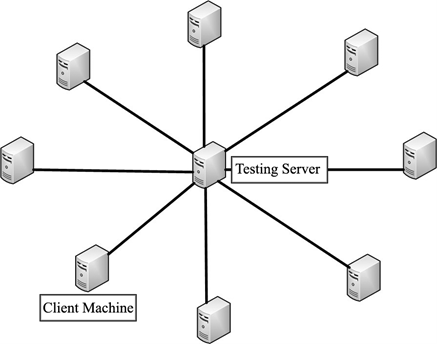}
         \caption{Hub and spoke layout for distributed demand-driven security testing}
    \label{fig:hubspoke}
\end{figure}

\subsection{Emerging Detection Methods}

The issues with the conventional methods in the main are that static analysis produces too many false positives, dynamic analysis does not scale, and code reviews are time consuming.  Research into newer methods tries to address these problems via some interesting and novel approaches.

\subsubsection{Distributed demand-driven security testing}

Proposed by \cite{ZHANG201460}, this involves many clients using OSS and one main testing server, in a hub and spoke style per Figure 1.  When a new path in a program is about to be exercised by user input, it is sent to the testing hub for security testing.  Symbolic execution is applied to the execution trace to check potential vulnerabilities on this new path, and if one is detected then a signature is generated and updated back to all the clients for protection.  If a path exercised by an input triggers any vulnerability that has already been detected, the execution is terminated.  This allows testing to focus on paths being used and helps stop attackers exploiting unreported vulnerabilities at a client site.

However, questions remain over how to handle large time and space overheads at client sites, how sensitive data is transmitted and handled, and actual implementation details are scarce.  That said, the principle of increasing test coverage of important paths as users exercise them is sound, and \cite{ZHANG201460} offers an interesting conclusion that machine learning can in future identify patterns of bugs at the testing server and use them to predict problematic code.

\subsubsection{Use of Execution Complexity Metrics}

\cite{Shin} examined complexity metrics collected during code execution, considering them potential indicators of vulnerable code locations.  Table I describes these metrics.  They measure the frequency of function calls and duration of execution functions using Callgrind, a Valgrind tool for profiling programs.  The collected data consists of the number of instructions executed on a run, their relationship to source lines, and call relationship among functions together with call counts.  Firefox and Wireshark were analysed with Callgrind to gather the metrics and results showed execution complexity metrics may be better indicators of vulnerable code than the conventional static complexity metric of lines of code, or LoC.   

\begin{table}[t]
\captionsetup{font=small}
\caption{Execution complexity metrics defined in \cite{Shin}}
	\label{tab:complexitymetrics}
	\scriptsize 
	\centering 
\begin{tabular}{|p{0.2\columnwidth}|p{0.5\columnwidth}|}
\hline
Name             & Definition                                    \\ \hline
NumCalls         & The number of calls to the functions defined in a file.                                                      \\ \hline
InclusiveExeTime & Execution time for the set of functions, $S$, defined in a file   including all the execution time spent by the functions called directly or   indirectly by the functions in $S$. \\ \hline
ExclusiveExeTime & Execution time for the set of functions, $S$,   defined in a file excluding the execution time spent by the functions called   by the functions in $S$.                            \\ \hline
\end{tabular}
\end{table}

Their initial results, shown in Table II, indicate the percentage of vulnerable files in execution is higher than the percentage of vulnerable files in total, and hence execution complexity metrics could be good indicators of vulnerability.  This can reduce the code inspection effort as prioritisation can take place based on the metrics.

\begin{table}[t]
\captionsetup{font=small}
\caption{Execution statistics from \cite{Shin}}
	\label{tab:execstats}
	\scriptsize 
	\centering 
\begin{tabular}{|p{0.1\columnwidth}|p{0.2\columnwidth}|p{0.3\columnwidth}|}
\hline
Program   & \% of vulnerable   files & \% of vulnerable   files in executed files \\
\hline
Firefox   & 3.8                      & 11                                         \\
\hline
Wireshark & 7.8                      & 19        
\\ \hline
\end{tabular}
\end{table}

\subsubsection{IDE Plugins for Early Detection}

\cite{SAMPAIO2016337} attempted to detect vulnerabilities earlier in the development process by using an Eclipse Java plug-in, arguing developers should be aware of security vulnerabilities as they are coding.  To reduce false positives, they proposed context-sensitive data flow analysis which uses a program’s context of variables and methods when searching for vulnerabilities instead of pattern matching.  \cite{ZHU2014449} presented interactive static analysis, also known as IDE static analysis.  They too developed an Eclipse Java plug-in for detecting code patterns that gives a two-way interaction between the IDE and the developer.  According to \cite{ZHU2014449}, their tool detected multiple zero day vulnerabilities.  Figure 2 shows a screenshot where the developer is instructed to annotate access control logic for a highlighted sensitive method call.

\begin{figure}[t]
    \centering
    \captionsetup{justification=centering}
    \captionsetup{font=small}
    \includegraphics[width=\columnwidth]{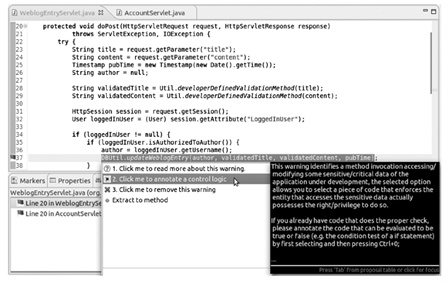}
         \caption{An IDE static analysis tool from \cite{ZHU2014449}}
    \label{fig:staticanalysis}
\end{figure}

\subsubsection{Machine Learning}

Machine learning is a type of artificial intelligence where computers use algorithms to learn iteratively, teaching themselves to recognise patterns.  Most OSS code is managed using version control systems like Git or CVS, with vulnerable code inserted via commits from the developer to the main data repository.  But many tools can’t run on a small code snippet in an individual commit, and checking the whole project is time consuming.  \cite{Perl} implemented a type of machine learning algorithm called a Support Vector Machine (SVM) that used metadata from commits made to OSS repositories.  The SVM used features from the metadata such as the number of added, deleted or modified functions and how often a contributor had contributed to a given project before.  Their results showed that false positives were reduced by over 99\% compared to those generated by a static analysis tool - to be exact, their SVM driven tool generated 36 false positives compared to 5,460 generated from the static analysis tool.  The goal of their work was to reduce the chance of vulnerabilities getting from a vulnerable commit into the fully deployed software.  \cite{Grieco} also developed a machine learning tool to predict vulnerabilities for large scale software like operating systems.  They took the popular Debian OS as an example, since it has 30,000 programs and 80,000 bug reports.  Clearly, code flaws can be hard to find manually in a code base of that size, so the application of machine learning is of interest.  Their classification results were not conclusive but nevertheless, as an initial study, they showed promise for large-scale vulnerability detection only using binary executables, an approach which does not appear to have been attempted elsewhere.

\subsubsection{Further Knowledge Formalisation and Linking Repositories}

\cite{ALQAHTANI2016153} discussed formalising knowledge representation to determine transitive dependencies in software.  The idea is the various vulnerability repositories that exist online like the NIST National Vulnerability Database (NVD), or the Common Weakness Enumeration (CWE) database can be linked and simultaneously used to find out if a project is indirectly dependent on vulnerable components.

\section{Conclusions \& Future Work}

The global use of OSS presents such a huge number of attack vectors that discovering novel techniques of vulnerability detection is an essential area of research.  Of the new methods mentioned in this paper, machine learning, early detection IDE plug-ins and linking repositories show much promise for future work.  Machine learning lends itself well to feature-rich OSS which speeds up classification of vulnerable code and reduces the time burden on development teams.  Early detection IDE plug-ins will help developers implementing OSS to grow and consolidate their secure coding knowledge.  Linking repositories ensures better value from the separate, unconnected datastores of vulnerabilities as they presently exist.  It may also be possible to use machine learning to levarage all these options in a modular system.  Improvements in OSS vulnerability detection might be quicker to realise than one would think – \cite{EnglishExton} mention Pareto’s law, where 80\% of effects can be contributed to 20\% of causes, and so identifying a small proportion of problematic OSS code then focusing testing efforts using a selection of detection methods could improve code quality and time-to-release, whilst reducing development and maintenance costs.  The exact mix of techniques might vary from one OSS scenario to another but in the first instance a strategy using a blend of methods that augment each other is likely to be considerably more performant than one approach in isolation.

\ifCLASSOPTIONcaptionsoff
  \newpage
\fi



%

 
\bibliographystyle{IEEEtran}
\bibliography{ccs-sample}

\end{document}